\documentclass[conference]{IEEEtran}

\usepackage[T1]{fontenc}
\usepackage{graphicx}
\usepackage{cite}
\usepackage{caption}
\usepackage{subcaption}
\usepackage{epstopdf}
\usepackage{algorithm}
\usepackage{algpseudocode}

\usepackage{overpic}
\usepackage{amssymb}
\usepackage{amsmath}
\usepackage{amsthm}
\usepackage{array}
\usepackage{color}
\usepackage{url}

\usepackage{overpic}
\usepackage{stfloats}
\usepackage{xspace}
\IEEEoverridecommandlockouts

\theoremstyle{plain}

\newtheorem{remark}{Remark}

\newcommand{\vect}[1]{\mathbf{#1}}

\def\diag{\mathrm{diag}}

\def\Htran{\mbox{\tiny $\mathrm{H}$}}
\def\Ttran{\mbox{\tiny $\mathrm{T}$}}
\def\imagunit{\mathsf{j}} 

\begin{document}

\title{Parametric Near-Field Channel Estimation for Extremely Large Aperture Arrays 
}

\author{
\IEEEauthorblockN{Alva Kosasih\IEEEauthorrefmark{1}, \"Ozlem Tu\u{g}fe Demir\IEEEauthorrefmark{2}, 
Emil Bj{\"o}rnson\IEEEauthorrefmark{1}
\thanks{This paper was supported by the Grant 2019-05068 from the Swedish Research Council.}}
\IEEEauthorblockA{\IEEEauthorrefmark{1}Department of Computer Science, KTH Royal Institute of Technology, Stockholm, Sweden (\{kosasih,emilbjo\}@kth.se)}
\IEEEauthorblockA{\IEEEauthorrefmark{2}Department of Electrical-Electronics Engineering, TOBB ETU, Ankara, Turkey  (ozlemtugfedemir@etu.edu.tr)}

}

\maketitle

\begin{abstract} 
Accurate channel estimation is critical to fully exploit the beamforming gains when communicating with extremely large aperture arrays. The propagation distances between the user and receiver, which potentially has thousands of antennas/elements, are such that they are located in the radiative near-field region of each other when considering the Fraunhofer distance of the entire array. Therefore, it is imperative to consider near-field effects to achieve proper channel estimation. This paper proposes a parametric multi-user near-field channel estimation algorithm based on  MUltiple SIgnal Classification (MUSIC) method to obtain the essential parameters describing the users' locations. We derive the estimated channel by incorporating the estimated parameters into the near-field channel model. Additionally, we implement a least-squares-based estimation corrector, resulting in a precise near-field channel estimation. Simulation results demonstrate that our proposed scheme outperforms classical least-squares and minimum mean-square error channel estimation methods in terms of normalized beamforming gain and normalized mean-square error.
\end{abstract}


\vspace{-2mm}

\IEEEpeerreviewmaketitle

\section{Introduction}

The realization of massive multiple-input multiple-output (M-MIMO) into 5G systems, across both sub-6 GHz and mm-wave bands, suggests that the next generation of wireless systems will likely exploit even larger arrays, referred to as the extremely large aperture array (ELAA) \cite{Bjornson2019d,2020_Wang_TWC,2018_Amiri_Globcomm}. Moreover, there is an ongoing trend toward employing higher frequencies implying a smaller wavelength in wireless systems  \cite{2019_Rappaport_Access,2020_Sanguinetti_JSAC}. As the array size increases while the wavelength shrinks, the Fraunhofer array distance which is the boundary between radiative near- and far-fields becomes large and thus the user equipments (UEs) are likely to fall into the radiative near-field region of the ELAA \cite{2020_Björnson_JCommSoc}. In the radiative near-field, a spherical curvature of the wavefront exists and therefore there are spherical phase variations over the antenna elements in the ELAA. The phase variations must be characterized by both angular and distance between the ELAA and the point source.  Consequently, conventional far-field codebooks, such as the discrete Fourier transform (DFT) codebook, are unsuitable due to non-orthogonality in the distance domain.

To address this issue, polar-domain representation for the extremely large-scale MIMO (XL-MIMO) channel has been proposed in \cite{2022_Cui_TC,2023_Wu_JSAC}. \cite{2022_Cui_TC} focuses on the recovery of angular and distance information in the near-field channel utilizing the sparsity in the polar domain, while \cite{2023_Wu_JSAC} considers the concept of location division multiple access (LDMA) to introduce additional spatial resolution in the distance domain. 
With the polar-domain representation, one can sample both angular and distance domains to obtain the near-field codebook, where the orthogonality  is asymptotically achieved with respect to the number of antennas. Moreover, the polar-domain representation is used to ensure a sparsity representation of the near-field channel enabling the utilization of compressive sensing methods such as the classical orthogonal matching pursuit (OMP) and simultaneous iterative gridless weighted (SIGW) algorithms \cite{2022_Cui_TC}. 
However, the orthogonality of the near-field codebook remains an issue for a practical number of antennas and in particular when a uniform planar array (UPA) is employed. This is because there exists a correlation between two adjacent grids \cite{2023_Guo_TVT}.

Another solution to estimate the near-field channel is by first estimating the parameters of the UEs' locations. Subsequently, the parameters are substituted into the channel parametric model to yield the channel estimate.
A straightforward approach is to discretize the three-dimensional ($3$D) spatial domain and perform classical spectral estimation methods (e.g., MUltiple SIgnal Classification (MUSIC)) to estimate the location of the UEs in the $3$D domain. However, such an approach is highly complex and performs badly when there are many UEs since the peaks may conflate.
In \cite{2012_He_TSP}, a two-step MUSIC algorithm was proposed, where the MUSIC is performed first in the angular domain and subsequently in the distance domain. It has been shown that the two-step MUSIC can achieve a high estimation performance and works for both near- and far- fields. However, the implementation was restricted in the case of uniform linear array (ULA) with a specific consideration of  sensor array models. 

In this paper, we leverage the two-step MUSIC approach to not only estimate the location of the UEs but also estimate their channels. This is facilitated by the parametric near-field channel model given in \cite{2020_Björnson_JCommSoc}. In contrast to \cite{2012_He_TSP}, we consider a UPA where the UEs are located in the near-field of the array. In addition, we refine our parametric channel model using a least-squares (LS) based estimation corrector resulting in an accurate near-field channel estimate. The simulation results show that our proposed method significantly outperforms the non-parametric LS and  regularized LS (R-LS) in terms of the normalized mean-square error (NMSE) and beamforming gain.

\section{System Model}

We consider a base station (BS) equipped with a UPA of $N$ identical aperture antennas, where $\sqrt{N}$ is an integer for simplicity, serving $K$ single-antenna user equipments (UEs). The location of UE $k$ is denoted as $(x_k,y_k,z_k)$, which is in the radiative near-field region (Fresnel region) of the BS array, for $k \in \{1,\dots,K\}$.  The BS antennas are deployed edge-to-edge on a square grid in the $xy$-plane. Let $n \in \{1,\ldots,\sqrt{N}\}$ and $m \in \{1,\ldots,\sqrt{N}\}$ denote  row and column indexes in the $y$ dimension, respectively. The vertical and horizontal inter-antenna distance is $D/\sqrt{2}$, which leads to the diagonal length of $D$ for each antenna. The antenna with index $(n,m)$ is then centered at the point $(\bar{x}_n,\bar{y}_m,0)$ given by
\begin{equation}
\bar{x}_n = \left(n-\frac{\sqrt{N}+1}{2} \right) \frac{D}{\sqrt{2}}, \,\, \bar{y}_m = \left(m-\frac{\sqrt{N}+1}{2} \right) \frac{D}{\sqrt{2}}
\end{equation}
and covers the area
\begin{align} 
\!\!\mathcal{A}_{n,m} = \left\{ (x,y,0) : \left| x - \bar{x}_n \right| \leq \frac{D}{\sqrt{8}},  \left| y - \bar{y}_m  \right| \leq \frac{D}{\sqrt{8}} \right\}.
\end{align}
For an impinging wave with $E_k(x,y)$ from UE $k$, the complex-valued channel  to receive antenna $(n,m)$ is \cite[Eqn. (6)]{2021_Björnson_Asilomar}
\begin{equation}
h_{n,m}^k = \frac{1}{E_{0,k}} \sqrt{\frac{2}{D^2}} \int_{\mathcal{A}_{n,m}}  E_k(x,y) dx \, dy
\end{equation} 
where $E_{0,k}$ is the electrical intensity of the radiated signal, 
\begin{equation} \label{eq:intensity-function}
E_k(x,y) = \frac{E_{0,k}}{\sqrt{4 \pi}} \frac{\sqrt{z_k \left(\left(x-x_k\right)^2+z_k^2\right)}} {r_k^{5/2}(x,y)} e^{-\imagunit \frac{2\pi}{\lambda}r_k(x,y)}, 
\end{equation}
and $r_k(x,y) =  \sqrt {\left(x-x_k\right)^2+(y-y_k)^2+z_k^2 }$ is the Euclidean distance between the array and UE $k$. 
For electrically small antennas (i.e., $D \ll \lambda$), where the UE is in the far-field of the BS's individual antenna elements, $E_k(x,y)$ is approximately constant over each antenna. Then, we can write 
\begin{equation}\label{eq:ch_resp}
  h_{n,m}^k(x_k,y_k,z_k) = \frac{D}{\sqrt{8 \pi}} \frac{\sqrt{z_k \left(\left(\bar{x}_n-x_k\right)^2+z_k^2\right)}}  {(r_{n,m}^k)^{5/2}}  e^{-\imagunit \frac{2\pi}{\lambda}r_{n,m}^k},
\end{equation}
where $r_{n,m}^k = \sqrt{\left(\bar{x}_n-x_k\right)^2+(\bar{y}_m-y_k)^2+z_k^2}$ is the Euclidean distance between the $(n,m)$-th antenna and UE $k$. 
Let us now define the near-field channel array response vector $\vect{a}(x_k,y_k,z_k)$  parameterized by the location of UE $k$ as
\begin{align}\label{ori_arr_resp}
    \vect{a}(x_k,y_k,z_k) = \left[h^k_{1,1}, \dots, h^k_{1,\sqrt{N}}, \dots,  h^k_{\sqrt{N},1}, \dots, h^k_{\sqrt{N},\sqrt{N}}  \right]^\mathrm{T}.
\end{align}
At time instance $l$, the received signal can be written as
\begin{equation}\label{eq:received_signal}
    \vect{q}[l] =  \vect{A} \vect{s}[l] + \vect{w}[l], \quad l=1,\ldots,L,
\end{equation}
where $\vect{A} = \left[ \vect{a}(x_1,y_1,z_1),   \dots, \vect{a}(x_K,y_K,z_K)\right]$, $\vect{q}[l] = [q_1[l],\dots,q_N[l]]^\mathrm{T}$ contains the received signals, $\vect{s}[l] = [{s_1}[l], \dots, {s}_K[l]]^\mathrm{T}$ is the pilot vector from $K$ UEs, and $\vect{w}[l] = [{w_1}[l], \dots, {w_N}[l]]^\mathrm{T}$ is additive noise where each entry follows and independent complex Gaussian distribution  with zero mean and variance $\sigma^2$. The noise is normalized such that the signal-to-noise ratio (SNR) is $ 1/\sigma^2$. 
We consider the case where the length of the pilot sequence is smaller than the number of UEs, $L  < K $.

\section{Near-Field Parametric Channel Estimation}

\begin{figure*}
\centering
\subfloat[$L=10$]
{\includegraphics[scale=0.5]{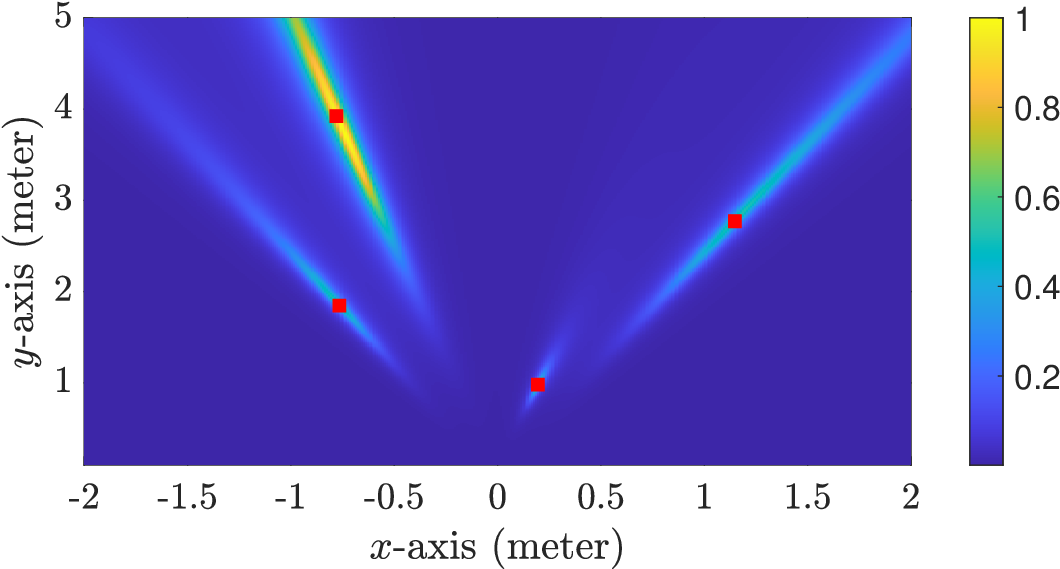}} \hfill
\centering
\subfloat[$L=3$]
{\includegraphics[scale=0.5]{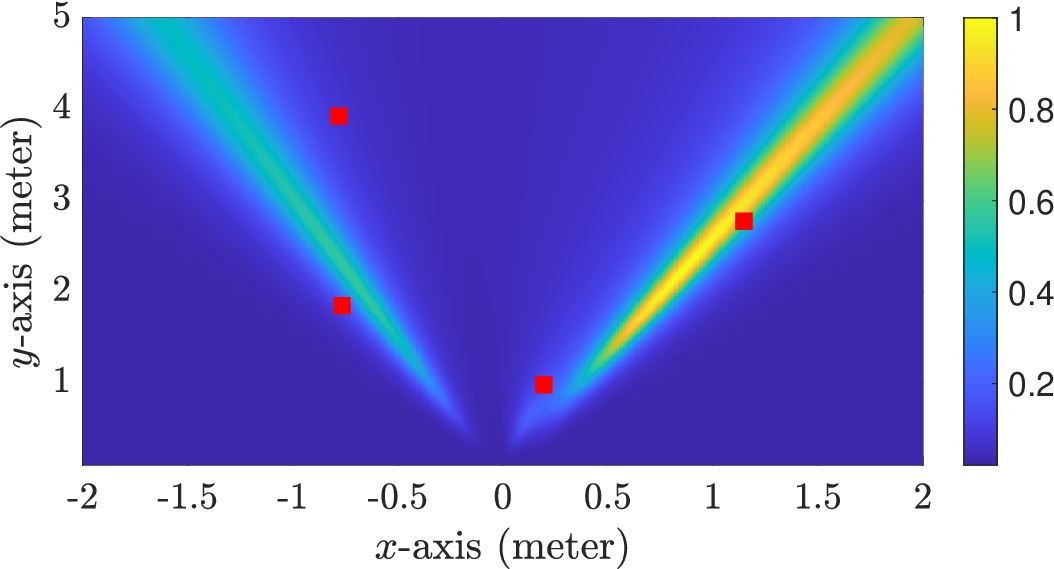}}
\caption{Parametric near-field channel estimation for large arrays using $3$D MUSIC with  $N=100, K=4$, 
${\rm SNR}=20$\,dB, $D =  \lambda/\sqrt{2}$, and  $\lambda = 0.1$\,m. 
The BS array is located at $(0,0,0)$ while the UEs are distributed across the $xy$-plane within the radiative near-field region of the BS array, indicated by red squares. }
\label{F_direct_MUSIC}
\end{figure*}
The channel response function, defined in \eqref{eq:ch_resp}, allows us to characterize the UEs' channels by estimating the locations $(x_k,y_k,z_k), k=1,\dots, K$, from $L$ transmissions at distinct instants. In the following, we provide a novel way to estimate the UEs' locations using the MUSIC method. We assume that: 1) The UEs are not located in exactly the same angular directions.  2) The pilot sequences are generated from independently and identically zero-mean complex Gaussian distribution, but are known by the BS and UEs. 3) The noise is independent of all the signals.

The MUSIC algorithm works by exploiting the structure of the eigenvectors in the sample covariance matrix:
\begin{align}
\widehat{\vect{R}}_L = \frac{1}{L} \sum_{l=1}^L \vect{q}[l]\vect{q}^{\Htran}[l].
\end{align}
Given the number of UEs $K$, we first construct  the noise-subspace matrix $\widehat{\vect{U}}_{\rm n}\in \mathbb{C}^{N \times (N-K)}$ whose columns are the eigenvectors of $\widehat{\vect{R}}_L$ corresponding to the smallest $(N-K)$ eigenvalues. Then, the $3$D MUSIC spectrum is generated as
\begin{align} \label{eq:MUSIC-spectrum}
S(x,y,z)=\frac{1}{\vect{a}^{\Htran}(x,y,z)\widehat{\vect{U}}_{\rm n}\widehat{\vect{U}}_{\rm n}^{\Htran}\vect{a}(x,y,z)},
\end{align}
where each possible value of $\vect{a}(x,y,z)$ is obtained by sampling the $3$D spatial domain using  \eqref{ori_arr_resp}. $K$ combinations of $(x,y,z)$ corresponding to the peaks in the MUSIC spectrum are then identified, each represents the UE's location.

Fig.~\ref{F_direct_MUSIC} illustrates the results of performing  MUSIC by searching for peaks in the spectrum expression in \eqref{eq:MUSIC-spectrum} across the $xyz$-dimensions. In this figure, we set the elevation angle to $0$ and therefore  the $3$D MUSIC degenerates to $2$D MUSIC. However, this assumption is not realistic because UEs can occupy any location in $3$D space. Furthermore, as shown in Fig.~\ref{F_direct_MUSIC}(b), we observe that when the number of pilot transmissions (referred to as the number of snapshots in MUSIC) is smaller than the number of UEs, MUSIC fails to yield peaks indicating the UEs' locations due to the spectral conflation among neighbors. We summarize the two problems when applying the $3$D MUSIC  algorithm:
\begin{itemize}
    \item \textit{Grid search problem.} The MUSIC algorithm needs to search the peaks by discretizing the spatial domain. When we consider the same number of grids in each dimension, the number of possible grids grows cubically for the $3$D MUSIC leading to an excessively high complexity problem.
    \item \textit{Number of snapshots problem.} The MUSIC performance with a low number of snapshots is unacceptable due to rank deficiency in the sample covariance matrix $\hat{\vect{R}}_L$. As illustrated in Fig.~\ref{F_direct_MUSIC}(b), when $L=3$, no distinct peaks can be detected to specify the UEs' locations. 
\end{itemize}

\subsection{Proposed Two-Step MUSIC}

To address the high complexity issue, we present a two-step MUSIC algorithm that operates through sequential estimations of both the angular and distance domains.
We first transform the channel response expression in \eqref{eq:ch_resp} into its polar domain equivalent, parameterizing by the azimuth angle $(\varphi)$, elevation angle $(\theta)$, and distance $(d)$. This transformation is defined as follows: $x_k = d_k \cos(\theta_k) \sin(\varphi_k)$, $y_k = d_k \sin(\theta_k)$, and $z_k = d_k \cos(\theta_k) \cos(\varphi_k)$. We then tightly approximate the phase variation in the channel response using the Fresnel approximation (first-order Taylor approximation), where we assume that the amplitude variations across the aperture are negligible in the considered radiative near-field, as discussed in \cite{2021_Björnson_Asilomar}. More specifically, we approximate the Euclidean distance between UE $k$ and $(n,m)$-th antenna element, as follows:
\begin{align}\label{eq:Fresnel_app}
  &  r_{n,m}^k \nonumber\\ \notag
    &= d_k\sqrt{1 + \frac{  \bar{x}_n^2 +\bar{y}_m^2 - 2 d_k (\cos(\theta_k) \sin(\varphi_k) \bar{x}_n + \sin(\theta_k)\bar{y}_m  )} {d_k^2}} \\ 
    &\approx  d_k + \frac{  \bar{x}_n^2 +\bar{y}_m^2}{2 d_k} - \cos(\theta_k) \sin(\varphi_k) \bar{x}_n - \sin(\theta_k)\bar{y}_m.  
\end{align}
The Euclidean distance determines the phase variation.
By treating all the terms involving $d_k$ as constants, 
we define a far-field array response vector, parameterized by ($\varphi,\theta$) for UE $k$ as 
\begin{align}\label{eq:app_arr_resp}
    \vect{\tilde{a}}(\varphi,\theta) = \bigg[\tilde{h}_{1,1}(\varphi,\theta), 
    \dots, \tilde{h}_{\sqrt{N},\sqrt{N}}(\varphi,\theta)  \bigg]^{\Ttran},
\end{align}
where $\tilde{h}_{n,m}(\varphi,\theta) = e^{j \frac{2\pi}{\lambda} \left(\cos(\theta) \sin(\varphi) \bar{x}_n + \sin(\theta)\bar{y}_m \right)}$.  
Using the far-field array response, we compute the $2$D MUSIC spectrum over the $2$D angular domain as 
\begin{align} \label{eq:2D_MUSIC-spectrum}
S(\varphi,\theta)=\frac{1}{\vect{\tilde{a}}^{\Htran}(\varphi,\theta)\widehat{\vect{U}}_{\rm n}\widehat{\vect{U}}_{\rm n}^{\Htran}\vect{\tilde{a}}(\varphi,\theta)}.
\end{align}
 Subsequently, we identify the $K$ tallest peaks corresponding to estimated azimuth and elevation angles, denoted as $\hat{\varphi}_k$ and $\hat{\theta}_k$, respectively, for $ k=1,\dots, K$. These peaks are defined as locally maximum points that are strictly greater than their immediate neighbors. We refer to the angular estimation as the first step of our proposed two-step MUSIC. 

The second step is to estimate the distance $d_k$ of UE $k$, given the estimated azimuth and elevation angles. 
More specifically, we substitute  $\hat{\varphi}_k$ and $\hat{\theta}_k$ into the polar domain transformation of the original array response, as defined by
\begin{equation}\label{ori_arr_resp_polar}
    \vect{a}(\hat{\varphi}_k,\hat{\theta}_k,d) = \bigg[h^k_{1,1}(\hat{\varphi}_k,\hat{\theta}_k,d) , 
    \dots, h^k_{\sqrt{N},\sqrt{N}}(\hat{\varphi}_k,\hat{\theta}_k,d)   \bigg]^{\Ttran},
\end{equation}
where $h^k_{n,m}(\hat{\varphi}_k,\hat{\theta}_k,d)$ is obtained from the polar domain representation of the original channel response in \eqref{eq:ch_resp} by neglecting the amplitude variations.
Subsequently, we calculate a $1$D MUSIC spectrum with respect to the distance domain as
\begin{align} \label{eq:1D_MUSIC-spectrum}
S(d)=\frac{1}{\vect{a}^{\Htran}(\hat{\varphi}_k,\hat{\theta}_k,d)\widehat{\vect{U}}_{\rm n}\widehat{\vect{U}}_{\rm n}^{\Htran}\vect{a}(\hat{\varphi}_k,\hat{\theta}_k,d)}.
\end{align}
\begin{remark}\label{Remark_2D}
Unlike the cubically growing complexity of the $3$D spectrum search in \eqref{eq:MUSIC-spectrum}, our proposed two-step MUSIC has a quadratic complexity in the first step and linear complexity in the second step. In a concrete numerical example, where we discretize the space into $100$ points per dimension, the original $3$D MUSIC algorithm necessitates searching over a million possible combinations. In contrast, our approach involves considering only $10100$ possible spectrum, which is 99 times fewer than the original $3$D MUSIC algorithm. 
\end{remark}

\begin{figure}
    \centering
    \includegraphics[scale = 0.55]{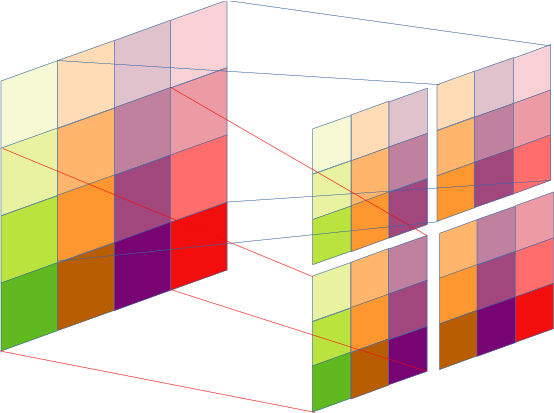}
    \caption{Illustration of additional snapshots from a dimensional reduction of the received signal. A UPA with $N=16$ is shown and we set $c_r = 1$.}
    \label{F_grouping_illustration}
\end{figure}

We have shown how we can reduce the complexity of performing MUSIC in near-field communications to estimate the angles as well as the distance by using two-step approach. However, the rank deficient sample covariance matrix remains to be an issue when only a few snapshots are available, as illustrated in Fig.~\ref{F_direct_MUSIC}(b). In what follows, we propose to create additional snapshots using spatial smoothing. For $K>1$, we reshape the received signal vector  into $\sqrt{N}\times\sqrt{N}$ matrix as
\begin{equation}
\vect{Q}=
\begin{bmatrix}
q_{1,1}[l] & \cdots & q_{1,\sqrt{N}}[l] \\
\vdots & \ddots & \vdots \\
q_{\sqrt{N},1}[l] & \cdots & q_{\sqrt{N},\sqrt{N}}[l]
\end{bmatrix}.
\end{equation}
Each element of the matrix is associated with the received signal in the antenna element of the UPA. We define a spatial smoothing parameter $c_r$, which determines the number of subarrays that will be used. For example, consider a UPA with a dimension of $\sqrt{N}\times\sqrt{N} = 4\times4$ and we set $c_r=1$. We will generate two sets of matrices: Two submatrices ($T =c_r+1 =2$) by horizontally overlapping the original matrix $\vect{Q}$, and another $ c_r+1$ times more by vertical overlap.
Each submatrix consists of $N_d = \sqrt{N}-c_r = 3$ elements in each row/column. This example is illustrated in Fig. \ref{F_grouping_illustration}. We denote the submatrix as $\vect{\tilde{Q}}_{(t_x,t_y)} \in \mathbb{C}^{N_d \times N_d}$, where $ t_x, t_y=1,\dots,T$. By vectorizing $\vect{\tilde{Q}}_{(t_x,t_y)} $, we obtain 
\begin{multline}
    \vect{\tilde{q}}_{(t_x,t_y)}[l]  = \big[ q_{t_x,t_y}[l],\dots, q_{t_x,t_y+N_d-1}[l],\dots \\
    q_{t_x+N_d-1,t_y}[l],\dots, q_{t_x+N_d-1,t_y+N_d-1}[l] \big]^{\Ttran}.
\end{multline}
We note that $N_d^2 \geq K$ is essential to maintain sufficient information for estimating the location of the UEs' locations. 
By utilizing the additional snapshots, we can calculate the sample covariance matrix as 
\begin{align}\label{eq:corr_group}
\widehat{\mathbf{R}}_{L,T} = \frac{1}{L T^2} \sum_{l=1}^L \sum_{t_x=1}^{T} \sum_{t_y=1}^{T}  \mathbf{\tilde{q}}_{t_x,t_y}[l] \mathbf{\tilde{q}}_{t_x,t_y}^{\Htran}[l].
\end{align}
Notice that with the additional snapshots, we will be able to satisfy the condition $LT^2 \geq K$ which is necessary to ensure sufficient rank of the sample covariance matrix.
Using the new sample covariance matrix, we calculate the far-field array response as defined in \eqref{eq:app_arr_resp}, followed by the computation of the $2$D MUSIC spectrum in \eqref{eq:2D_MUSIC-spectrum}, and finally, the $1$D MUSIC spectrum as in \eqref{eq:1D_MUSIC-spectrum}. 
Since the dimension of the sample covariance matrix is $N_d \times N_d$, where $N_d = \sqrt{N}-c_r$, the indices ${n}$ and ${m}$ are within the range ${1,\ldots,N_d}$ when computing equations \eqref{eq:app_arr_resp} through \eqref{eq:1D_MUSIC-spectrum}.
Once we have acquired the estimated parameters, $(\hat{\varphi}_k, \hat{\theta}_k, \hat{d_k})$, we convert these parameters back to the Cartesian domain $(\hat{x}_k, \hat{y}_k, \hat{z}_k)$ and utilize them to compute the channel estimate of UE $k$, for $k=1,\dots,K$, as in  \eqref{eq:ch_resp}.

 \subsection{Estimation Correction}

The estimation errors of the UEs' locations will result in channel estimation errors in the parametric model \eqref{eq:ch_resp}. 
To minimize the estimation errors, we propose refining the parametric channel estimate by scaling UE $k$'s channel vector estimate with a constant factor, denoted as $\alpha_k$. Through simulation, we discovered that the phase term in \eqref{eq:ch_resp} exhibited sensitivity to errors in the location parameters. 
Hence, $\alpha_k$ can be considered a phase correction for the channel estimation error, even though it also contributes to correcting the amplitude error.
We rewrite \eqref{eq:received_signal} to include the estimation correctors as 
\begin{equation}\label{eq:phase_err_model}\small
    \underbrace{\begin{bmatrix}
            \hat{h}_{1,1}^1 & \ldots & \hat{h}_{1,1}^K \\ \vdots & \ddots & \vdots \\ \hat{h}_{\sqrt{N},\sqrt{N}}^1 & \ldots & \hat{h}_{\sqrt{N},\sqrt{N}}^K
    \end{bmatrix}}_{\vect{\hat{A}}}
\underbrace{\begin{bmatrix}
s_1[l]& \ldots & 0 \\
\vdots & \ddots & \vdots \\
0 & \ldots &   s_K[l]
\end{bmatrix}}_{\diag{({\vect{s}}[l]})} 
    \underbrace{\begin{bmatrix}
         \alpha_1    \\   \vdots \\  \alpha_K  
    \end{bmatrix}}_{{\boldsymbol{\alpha}}}  +
    \underbrace{\begin{bmatrix}
            w_1[l] \\  \vdots \\   w_N[l]
    \end{bmatrix}}_{\vect{w}[l]}, 
\end{equation}
where ${\vect{\hat{A}}}\in \mathbb{C}^{N \times K}$ represents the channel matrix estimate, which is obtained by first converting  $(\hat{\varphi}_k,\hat{\theta}_k,\hat{d}_k)$ into the Cartesian domain to derive the UE's location estimate $(\hat{x}_k,\hat{y}_k,\hat{z}_k)$ and subsequently  substituting  it into the parametric model in \eqref{eq:ch_resp}. Hence, $\hat{h}_{n,m}^k$ represents the estimation of ${h}_{n,m}^k$ before applying the estimation correction.
We can expand the model for $L$ transmissions by  stacking   \eqref{eq:phase_err_model} as
\begin{equation}\label{eq:stacked_estimated_model}
    \underbrace{\left[\vect{I}_{L} \otimes \vect{\hat{A}} \right] 
    \begin{bmatrix}
        \diag{({\vect{s}}[1])} \\
        \vdots \\
        \diag{({\vect{s}}[L])}
    \end{bmatrix}}_{\vect{A}_{\rm Ds}\in \mathbb{C}^{LN\times K}}
    \boldsymbol{\alpha} + \vect{w},
\end{equation}
where  $\vect{w} = [\vect{w}^{\Ttran}[1],\dots,\vect{w}^{\Ttran}[L] ]^{\Ttran}\in\mathbb{C}^{LN}$ and $\vect{s}[l] = {[{s_1}[l], \dots, {s}_K[l]]}^{\Ttran}$.
Based on \eqref{eq:stacked_estimated_model}, we obtain the estimation of $\boldsymbol{\alpha}$ using the LS method as
\begin{equation}\label{eq:phase_corrector_estimate}
     \boldsymbol{{\hat{\alpha}}}  = (\vect{A}_{\rm Ds}^{\Htran} \vect{A}_{\rm Ds})^{-1} \vect{A}^{\Htran}_{\rm Ds} \vect{q}_{\rm R},
\end{equation}
where $\vect{q}_{\rm R} = [\vect{q}^{\Ttran}[1],\dots, \vect{q}^{\Ttran}[L]]^{\Ttran}$.
We summarize our proposed method to estimate the near-field channel using parametric estimation techniques in Algorithm~\ref{A1}.

\begin{algorithm}\small
\caption{The Near-Field Parametric Channel Estimation}
\label{A1}
\begin{algorithmic}[1]
\State {\textbf{Input: }$\vect{q}_{\rm R}, c_r$}
\State {Calculate the sample covariance matrix $\widehat{\vect{R}}_{L,T}$ in  \eqref{eq:corr_group}. }
    \Statex \textbf{Two-Step MUSIC:}   
    \State Compute the noise subspace $\widehat{\vect{U}}_{\rm n}$ based on $\widehat{\vect{R}}_{L,T}$.
    \For {$k=1,\dots, K$} 
        \State  Calculate the MUSIC spectrum in \eqref{eq:2D_MUSIC-spectrum} based on $\hat{\theta}_k$.
        \State Find the peak and assign the corresponding  $(\theta,\varphi)$ as  $(\hat{\theta}_k,\hat{\varphi}_k)$. 
        \State  Calculate the MUSIC spectrum  in \eqref{eq:1D_MUSIC-spectrum} based on $(\hat{\varphi}_k,\hat{\theta}_k)$.
        \State Find the peak and assign the corresponding $d$ as $\hat{d}_k$. 
    \EndFor    
    \State Transform $(\hat{\varphi}_k,\hat{\theta}_k,\hat{d}_k)$ into  $(\hat{x}_k,\hat{y}_k,\hat{z}_k)$ and substitute it into \eqref{eq:ch_resp} to obtain $\vect{\hat{A}}$.
    \Statex \textbf{Estimation Correction:}   
    \State Compute \eqref{eq:phase_corrector_estimate} to calculate the corrector vector $\boldsymbol{\alpha}$.
    \State The final channel matrix estimate is $\vect{\hat{A}} \diag{(\boldsymbol{\alpha})}$.
\State {\textbf{Output:  $\vect{\hat{A}} \leftarrow \vect{\hat{A}} \diag{(\boldsymbol{\alpha})}$ }  }  
\end{algorithmic}
\end{algorithm}

\section{Numerical Results}

In this section, we compare our proposed method with the classical non-parametric channel estimation methods, i.e., the LS and R-LS channel estimators. The LS is based on the pseudo-inverse of the pilot signal matrix $\vect{S} = [\vect{s}[1], \dots, \vect{s}[L]]$ while the R-LS channel estimate is expressed as
\begin{equation}
    \hat{\vect{A}}_{\rm RLS} = \vect{Q}_{\rm R} \left(\vect{S}^{\Htran} \vect{S}  + \sigma^2 \vect{I}_{{L} }\right)^{-1} \vect{S}^{\Htran},
\end{equation}
where $\vect{Q}_{\rm R} = [\vect{q}[1],\dots, \vect{q}[L]]$.
We consider a near-field communication system with $K=4$ UEs and a BS equipped with $N=100$ antennas deployed as a square array. The spacing between the antenna elements is $\lambda/2$, the frequency is  $3$\,GHz, and therefore $\lambda=0.1$\,m. The UEs are randomly placed within the near-field distance of the array, i.e., $[d_B,d_{FA}]$ (see \cite{2020_Björnson_JCommSoc}), where $d_B = 2 D \sqrt{N} = 1.4$\,m and $d_{FA} = 10$\,m. Furthermore, we consider random azimuth and elevation angles within the ranges of $[-4\pi/9, 4\pi/9]$ and $[-\pi/3, \pi/3]$, respectively. We ensure that UEs are well separated by $\pi/100$ radian in elevation/azimuth angular direction. 
\begin{figure}
\centering
\subfloat[Beamforming gain with respect to SNR with $c_r=1$.]
{\includegraphics[width=0.42\textwidth]{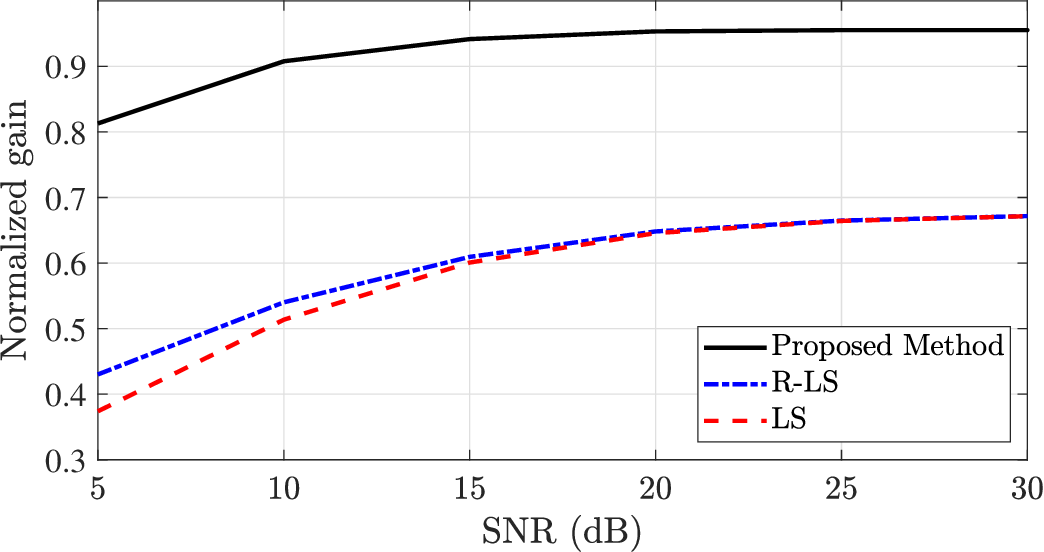}}\hfill
\subfloat[NMSE with respect to SNR with $c_r=1$.]
{\includegraphics[width=0.42\textwidth]{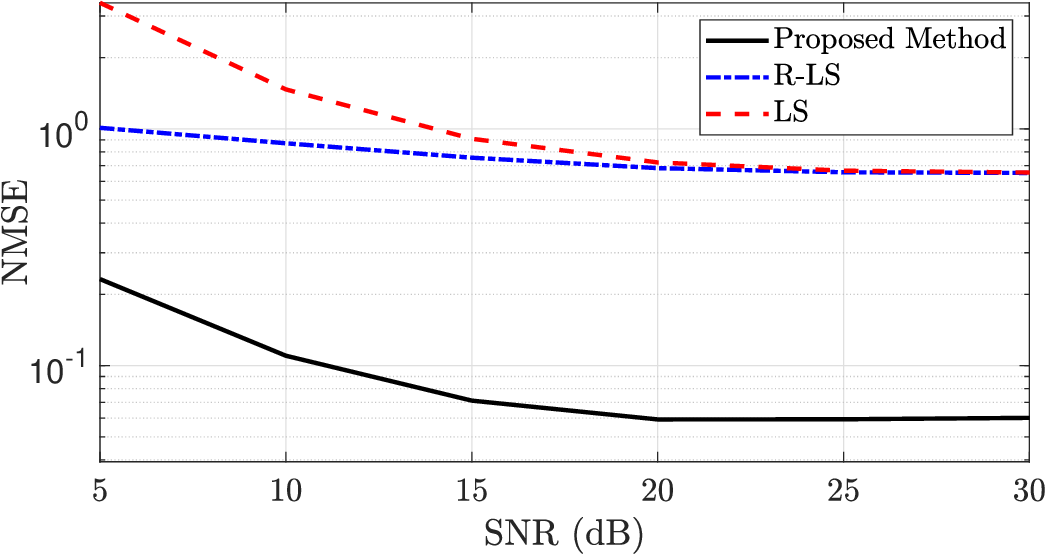}}\hfill
\caption{Performance comparison of the proposed scheme with classical channel estimation techniques.}
\label{F_N_10}
\end{figure}
In Fig.~\ref{F_N_10}(a), we illustrate how our proposed method outperforms LS and R-LS channel estimation techniques in terms of normalized beamforming gain. The beamforming gain for UE $k$ is defined as $ \mathbb{E}\left[ \left|\vect{v}_{\rm G}^{\Htran} \cdot \frac{  \vect{a}(x_k,y_k,z_k)  }{ \| \vect{a}(x_k,y_k,z_k)  \|}\right|^2 \right]$ , where ${\vect{v}_{\rm G}} = \frac{\hat{\vect{a}}_k}{\|\hat{\vect{a}}_k\|}$ and $\hat{\vect{a}}_k$ is the $k$-th column in the estimated channel matrix $\hat{\vect{A}}$.
We calculated the average of the normalized beamforming gain across all UEs, considering $10\,000$ channel realizations. The classical methods did not perform well, primarily due to the number of pilot transmissions being less than the number of UEs, which resulted in a {critical rank deficiency} in the matrices involved in the computation. We note that evaluating the performance of $3$D MUSIC is not possible due to the absence of distinct peaks in the considered simulation setup, as illustrated in Fig.~\ref{F_direct_MUSIC}(b).

In Fig.~\ref{F_N_10}(b), we show that the proposed method outperforms the classical LS and R-LS channel estimation methods in terms of NMSE. The NMSE for UE $k$ is defined as $\frac{ \mathbb{E}[ \|\vect{a}(x_k,y_k,z_k) - \hat{\vect{a}}_k\|^2 ]}{\mathbb{E}[\|\vect{a}(x_k,y_k,z_k)\|^2 ]}$. Similarly to the calculation of the beamforming gain, we average the NMSE over the UEs and across $10\,000$ channel realizations. Low NMSE values confirm the effectiveness of our proposed phase correction method, signifying both accurate phase alignment and proper scaling of the channel estimate with the actual channel.

\section{Conclusion}

We considered multi-user parametric near-field channel estimation with limited signaling. We proposed a two-step MUSIC algorithm to estimate the DoA and used this information to obtain the UEs' distances from the BS array. Our simulation results demonstrated that the proposed method significantly outperformed the classical channel estimation techniques.

\bibliographystyle{IEEEtran}

\bibliography{IEEEabrv,refs}

\end{document}